\documentclass[12pt,doublesp]{iopart}
\usepackage{iopams}
\usepackage{graphicx}
\begin{document}

\def \etcl {\mbox{(BEDT-TTF)$_{\rm 3}$Cl$_{\rm 2}\cdot$2H$_{\rm 2}$O}}
\def \cuscn{$\kappa$-(BEDT-TTF)$_2$Cu(NCS)$_2$}
\def \MR{magnetoresistance}
\def \fs{Fermi surface}
\def \Pc{$P_{\rm c}$}
\def \Bc2{$B_{\rm c2}$}
\def \deg{$^\circ$}
\def \thetat{$\theta_{\rm true}$}
\def \phit{$\phi_{\rm true}$}
\def \Bcperp{$B_{{\rm c2}\perp}$}
\def \Bcpara{$B_{{\rm c2}\parallel}$}
\def \pf6{(TMTSF)$_2$PF$_6$}
\def \Ef{$E_{\rm F}$}
\def \Ec{$E_{\rm c}$}

\def \trueangles{
\begin{eqnarray}
\cos\theta_{\rm
true}=\sin\theta\sin\epsilon\cos(\phi-\psi)+\cos\theta\cos\epsilon
\label{cost}
\\
\nonumber\\
\cos\phi_{\rm
true}=\frac{\sin\theta\cos\epsilon\cos(\phi-\psi)-\cos\theta\sin\epsilon}{\sin\theta_{\rm
true}} \\
\nonumber\\
\sin\phi_{\rm
true}=\frac{\sin\theta\sin(\phi-\psi)}{\sin\theta_{\rm true}}
\end{eqnarray}
}

\def \thetatrue90{
\begin{equation}
\cot\theta=-\tan\epsilon\cos(\phi-\psi)
\end{equation}
}

\def \GL{
\begin{equation}
B_{\rm c2}(\theta_{\rm true})=\frac{B_{{\rm
c2}\perp}}{\sqrt{\cos^2(\theta_{\rm
true})+\gamma^{-2}\sin^2(\theta_{\rm true})}}, \label{eqnGL}
\end{equation}
}

\def \coh{
\begin{equation}
B_{\rm c2\perp}=\frac{\Phi_0}{2\pi\xi_\parallel^2}, \label{eqncoh}
\end{equation}
}

\def \local{
\begin{eqnarray}
\lo{\sigma \propto}
\exp\left[\left(\frac{T_0}{T}\right)^{1/2}\left\{1-\left(1-\frac{1}{2}g\mu_{\rm
B}B/(E_{\rm c}-E_{\rm F})\right)^{\beta
d/n}\right\}\right]\nonumber\\
+\exp\left[\left(\frac{T_0}{T}\right)^{1/2}\left\{1-\left(1+\frac{1}{2}g\mu_{\rm
B}B/(E_{\rm c}-E_{\rm F})\right)^{\beta d/n}\right\}\right].
\label{eqnlocal}
\end{eqnarray}
}

\def \fgphase{
\begin{figure}[t]
\centering
\includegraphics[height=8.5cm]{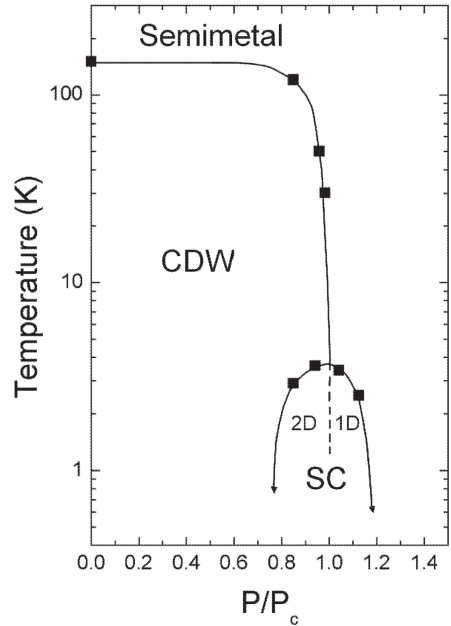}
\caption{The phase diagram of \etcl ~proposed by Lubczynski {\it
et al.}~\protect\cite{lub}. $P_{\rm c}$ is the critical pressure
at which the onset of the CDW state tends to zero temperature and
the pressure that separates the region of saturating \MR ~from the
region of diverging \MR ~in fields of up to 15~T as shown by the
dotted line. In Reference~\protect\cite{lub} \Pc~is found to be
close to 12~kbar.} \label{phase}
\end{figure}
}

\def \fgstruc{
\begin{figure}[t]
\centering
\includegraphics[height=8cm]{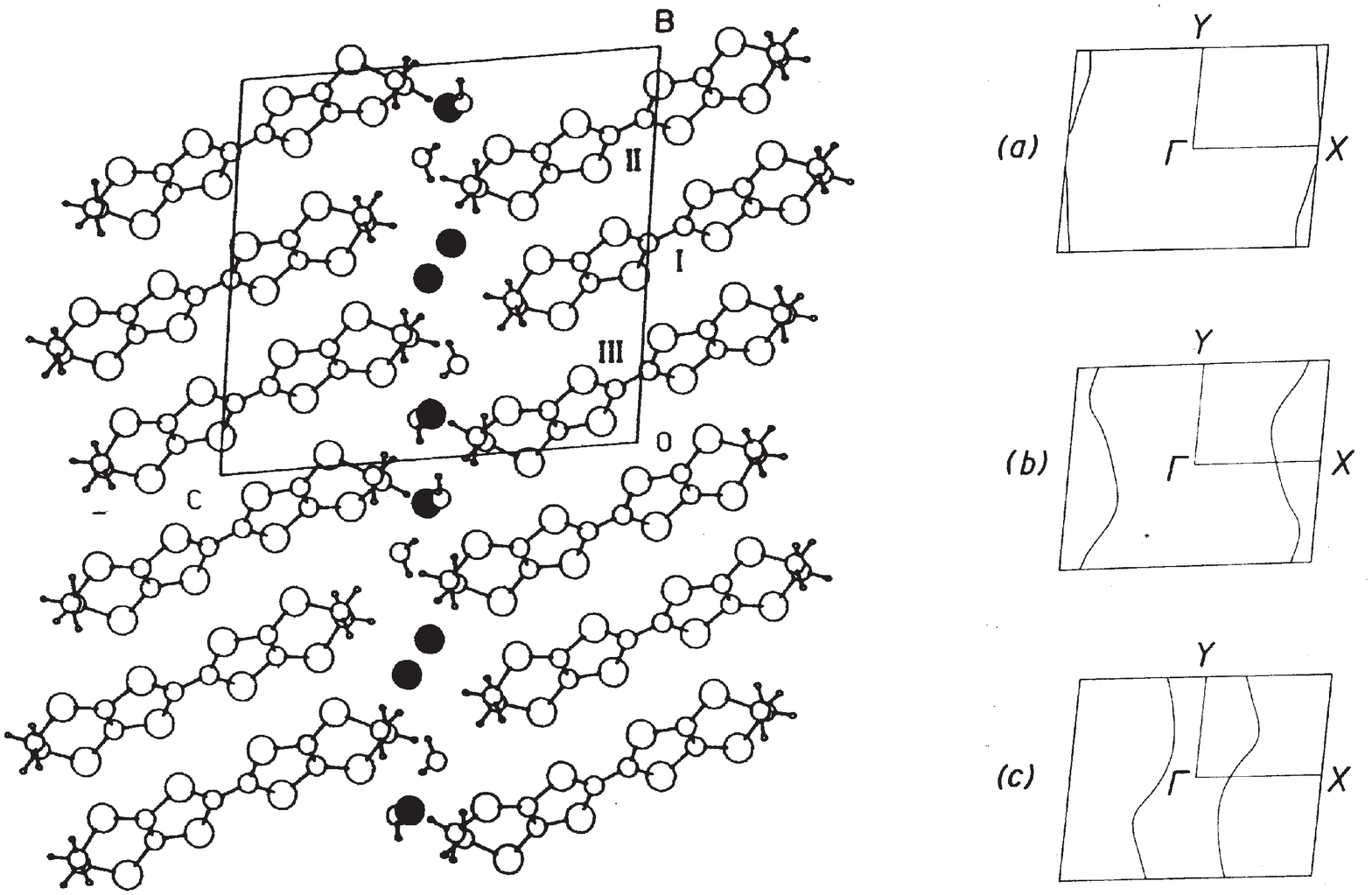}
\caption{Left: View of the unit cell of \etcl ~along the a-axis.
It consists of three independent (BEDT-TTF) molecules, two
chlorine atoms and two water molecules. Six (BEDT-TTF) molecules
collectively donate four electrons to each anion leaving each
molecule with an average charge of $+\frac{2}{3}$e
(from~\protect\cite{matt}). Right: The \fs ~of Whangbo {\it et
al.}~\protect\cite{whang}; (a) shows the Q2D closed electron
pocket, (b) the Q1D electron sheet and (c) the Q1D hole sheet.}
\label{struc}
\end{figure}
}

\def \figDAC{
\begin{figure}[t]
\centering \vspace{6cm} \includegraphics{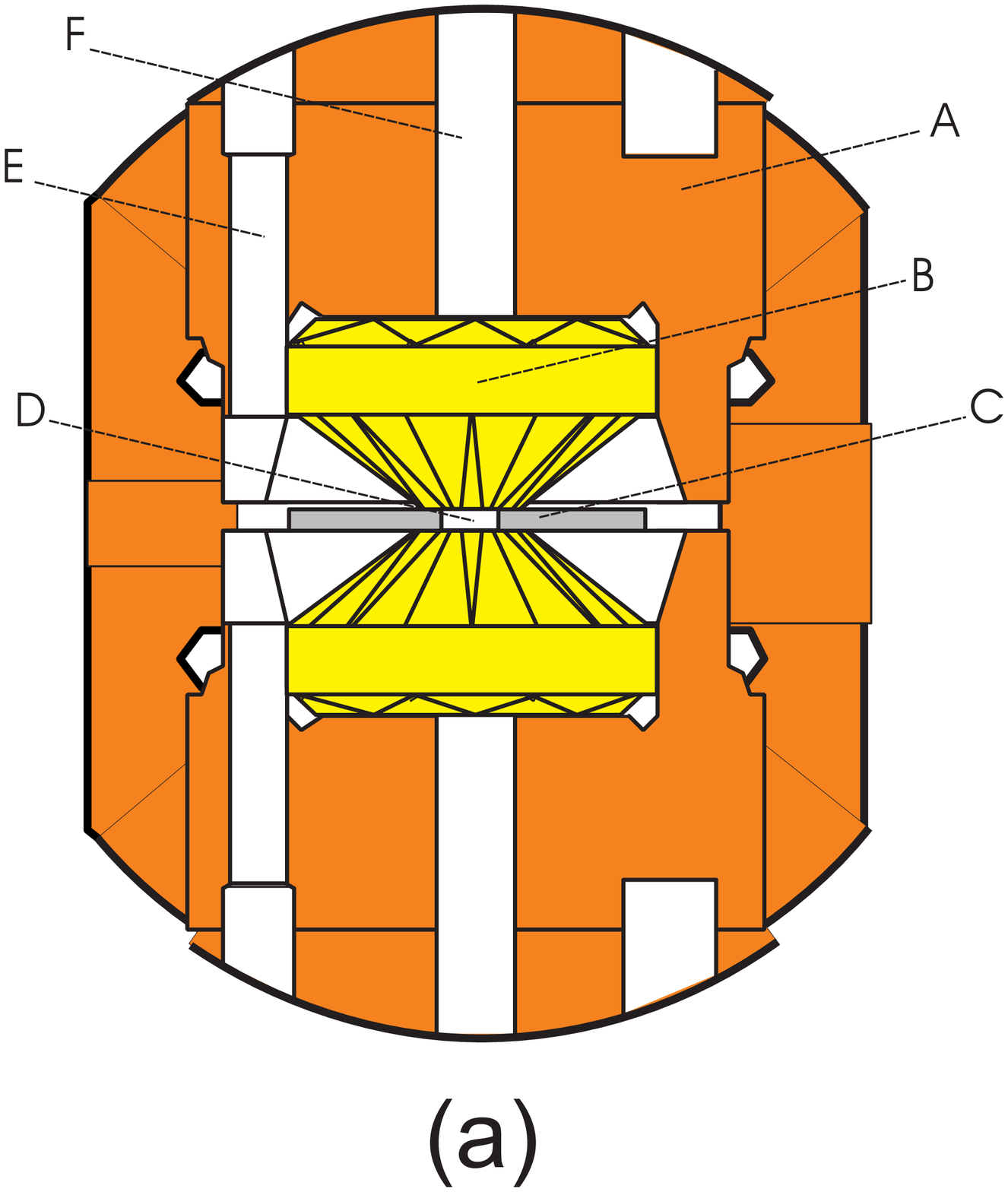} \includegraphics{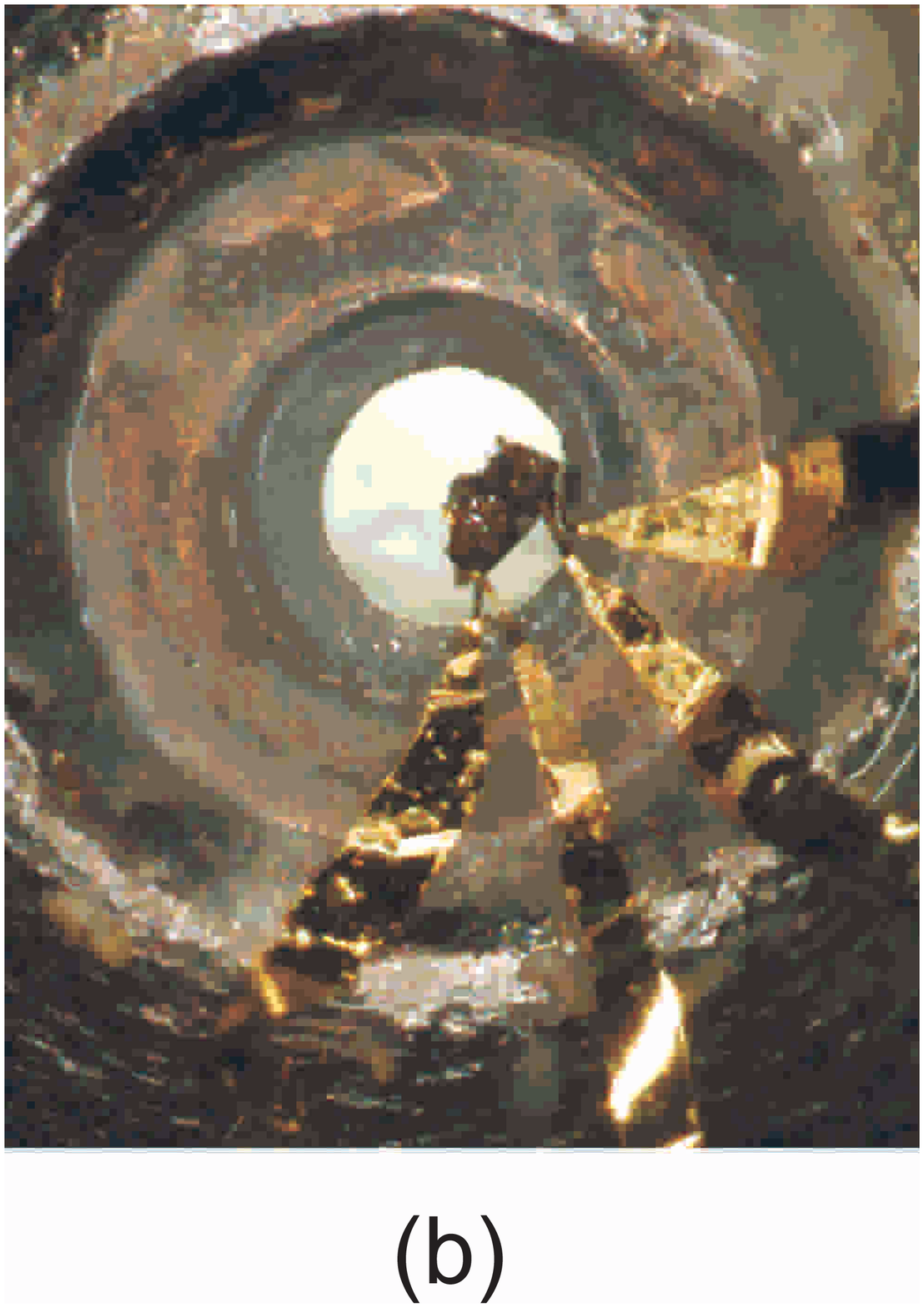} \caption{(a) Diagram
of the turnbuckle DAC with outer diameter 6.4~mm. A: cell body, B:
upper diamond, C: stainless steel gasket, D: sample cavity, E:
feedthrough for sample wires, F: threaded end cap, also used to
deliver radiation for exciting the ruby fluorescence used in
calibrating the pressure. (b) The \etcl~sample in position in the
gasket cavity prior to assembly. The cavity diameter is
$350~\mu$m.} \label{DAC}
\end{figure}
}

\def \fgaxis{
\begin{figure}[t]
\centering
\includegraphics[height=7cm]{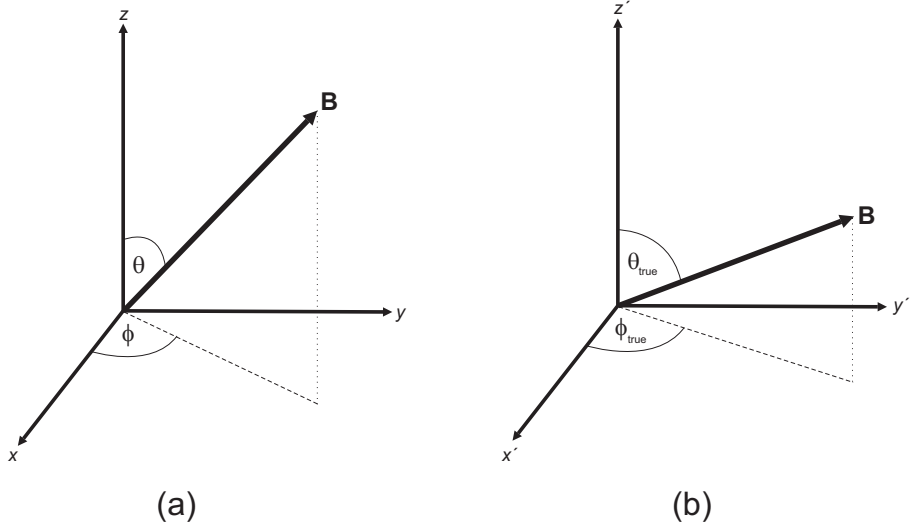}
\caption{(a) The laboratory frame, showing the relation between
$\theta$, $\phi$ and the magnetic field, ${\bf B}$. The $z$-axis
corresponds to the long axis of the DAC. (b) The sample frame,
showing the relation between \thetat, \phit~and the magnetic
field. The $z'$-axis corresponds to the normal to the
$ab$-planes.} \label{fgaxis}
\end{figure}
}

\def \figBc2{
\begin{figure}[t]
\centering
\includegraphics[height=8cm]{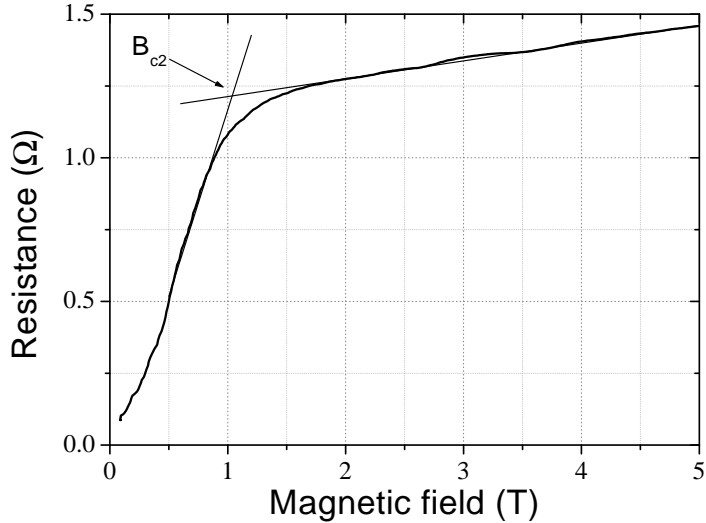}
\caption{Resistance of a single crystal of \etcl~as function of
magnetic field at 14.0~kbar, $T=0.5$~K and $\theta_{\rm
true}=171$\deg. The intersection of the straight lines defines
$B_{c2}$.} \label{figBc2}
\end{figure}
}

\def \figGL{
\begin{figure}[t]
\centering
\includegraphics[height=9cm]{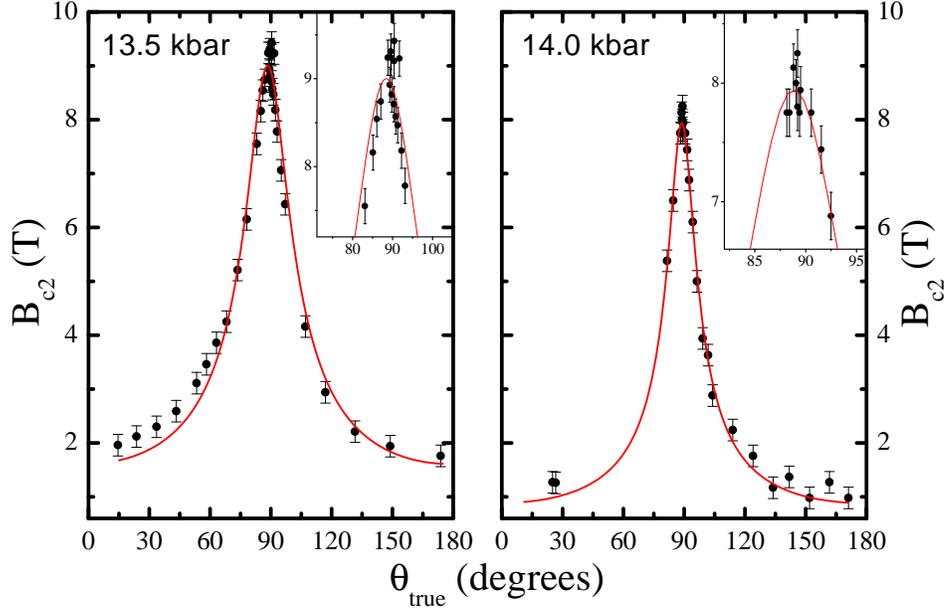}
\caption{The angular dependence of the upper critical field of
\etcl~at 13.5 kbar and 14.0 kbar. The data was taken at several
different values of \phit. The solid line is a fit to
equation~\ref{eqnGL} (see text). \thetat~is the angle between the
magnetic field and the normal to the conducting planes.}
\label{figGL}
\end{figure}
}

\def \tabBcdata{
\Table{\label{Bcdata}Results from the data in figure~\ref{figGL}.
\Bcperp~and $\xi_\parallel$~are found by fitting to
Equation~\ref{eqnGL}~and $B_{\rm PPL}$~is the Pauli paramagnetic
limit.} \br
&13.5 kbar&14.0 kbar\\
\mr
$T_{\rm c}$&$3.2\pm0.5$~K&$2.8\pm0.5$~K\\
$B_{\rm c2max}$&$9.2\pm0.2$~T&$8.3\pm0.2$~T\\
\Bcperp&$1.61\pm0.2$~T&$0.88\pm0.2$~T\\
$\xi_\parallel$&$143\pm18$~\AA&$193\pm44$~\AA\\
$B_{\rm PPL}=1.84T_{\rm c}$&$5.9\pm0.9$~T&$5.2\pm0.9$~T\\
\br
\endTable
}

\def \figAMRO{
\begin{figure}[t]
\vspace{9.5cm} \includegraphics{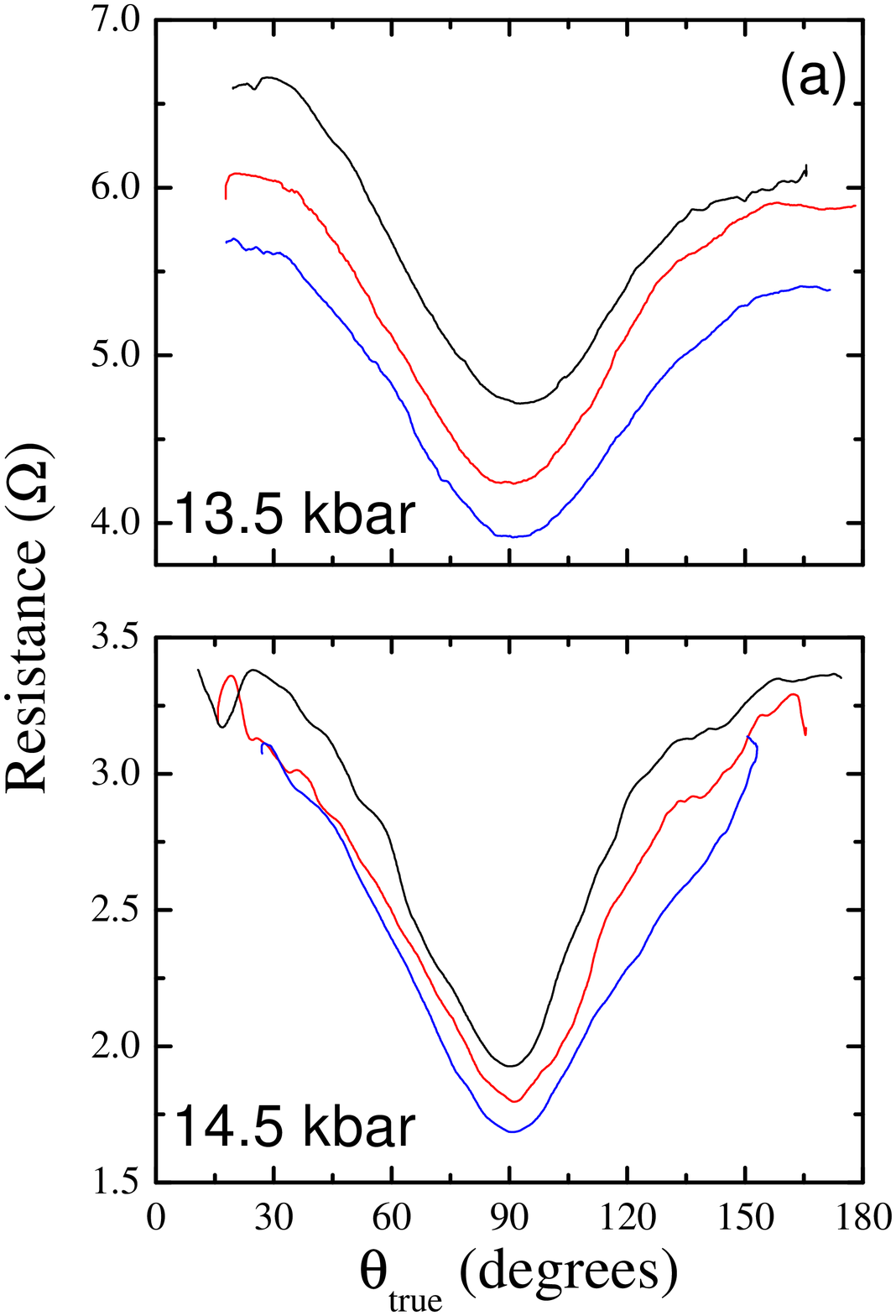} \includegraphics{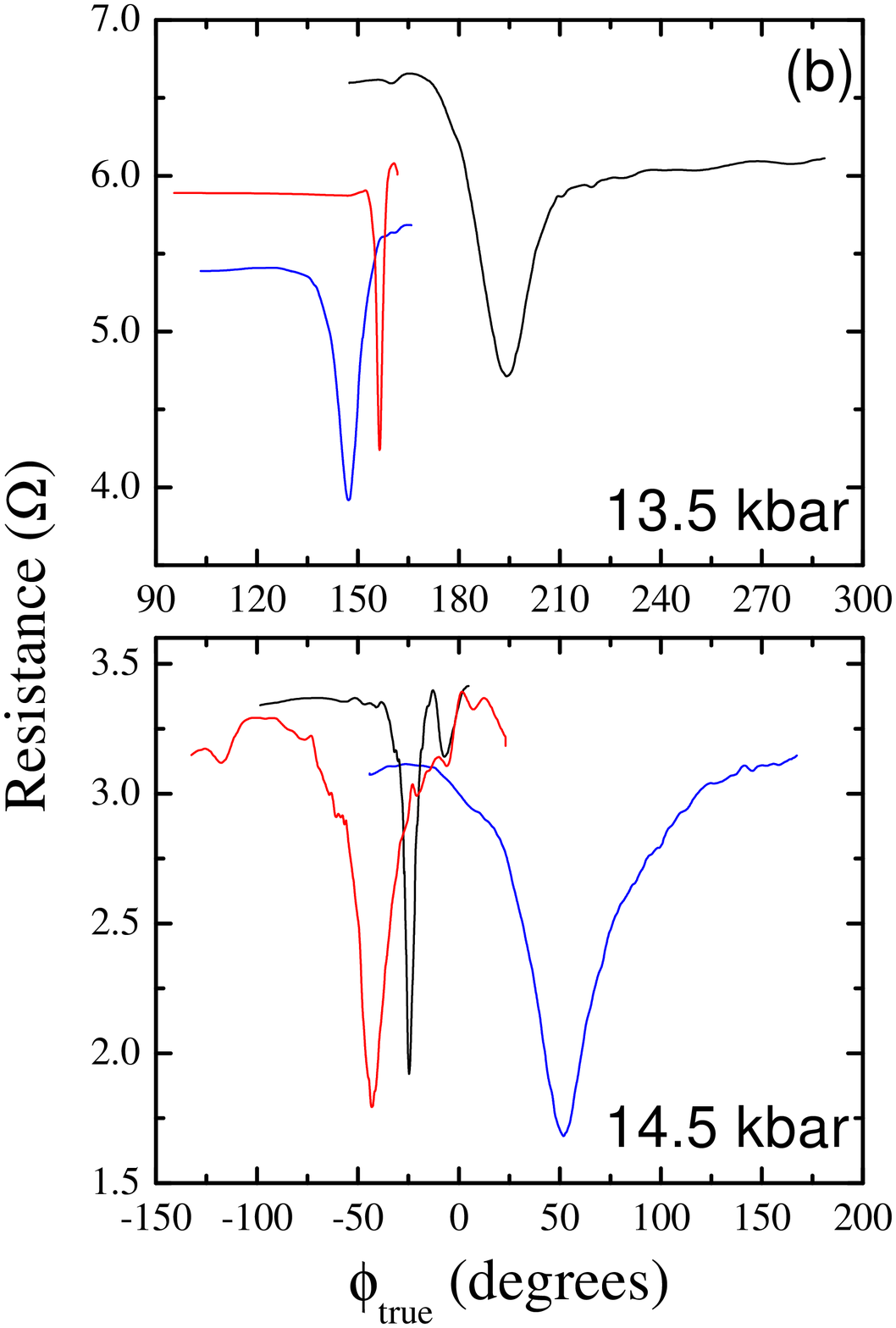} \nonumber
\vspace{10cm}\includegraphics{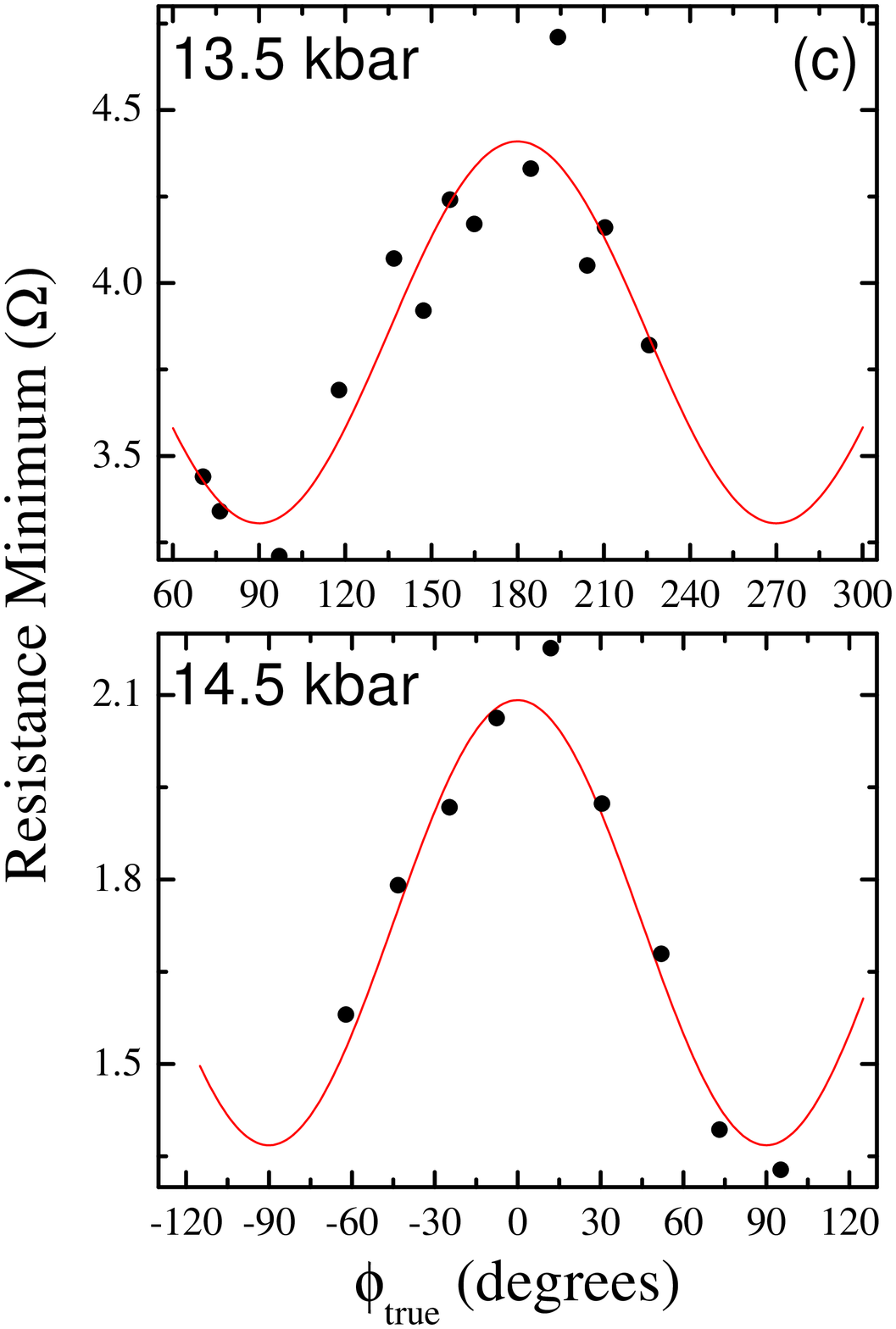} \caption{(a) Interlayer magnetoresistance
of \etcl~as a function of \thetat~with $B=30$~T and $T=0.5$~K at
13.5~kbar and 14.0~kbar. The three different curves are from three
different values of the measured $\phi$-angle. (b) The same curves
as in (a) shown as a function of \phit. (c) The points represent
the resistance minima at $\theta_{\rm true}=90$\deg~as a function
of \phit. The solid curves are fits to a single cosine function
with two-fold symmetry.} \label{AMRO}
\end{figure}
}

\def \fgMR13kb{
\begin{figure}[t]
\centering
\includegraphics[height=9cm]{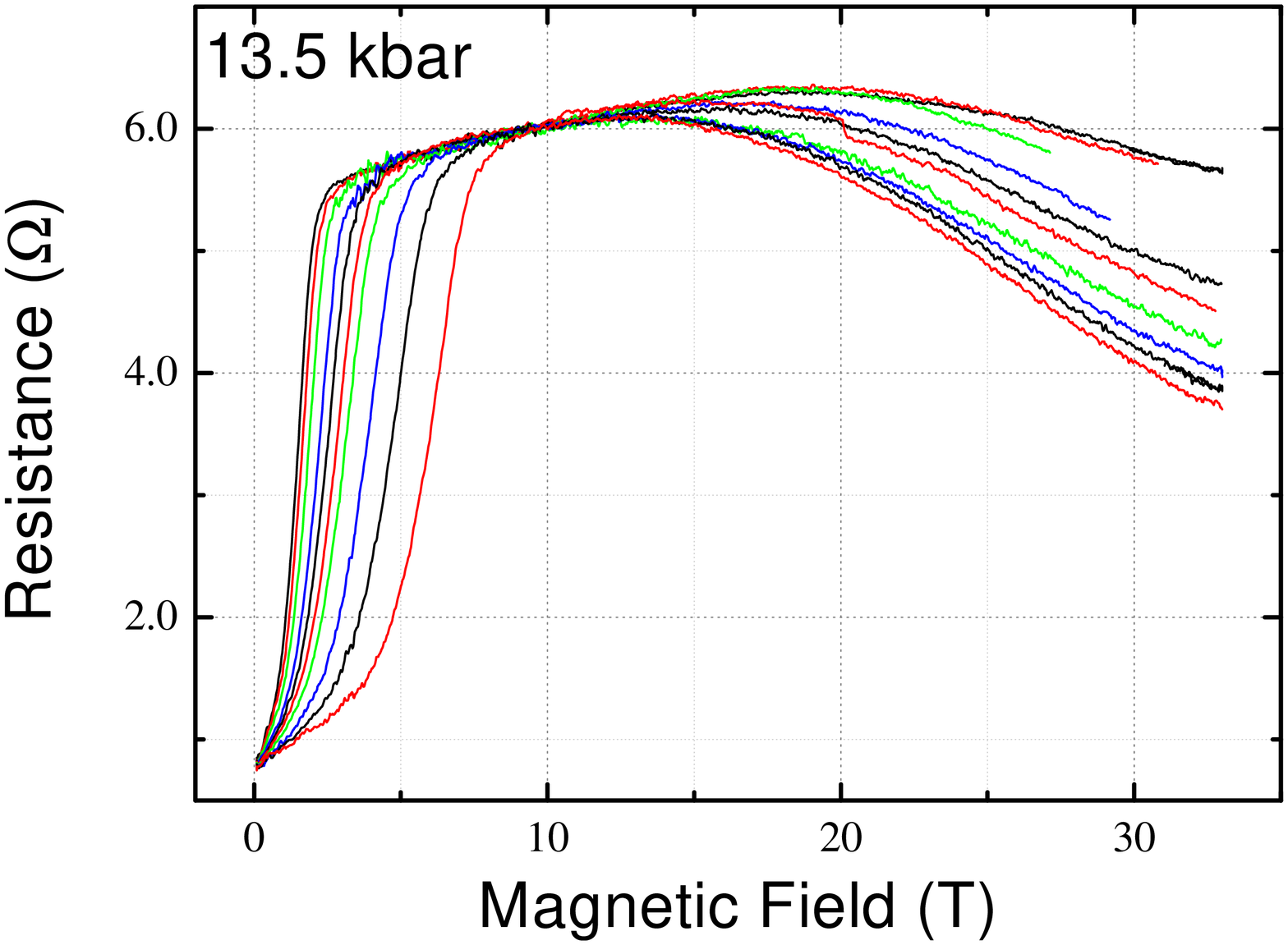}
\caption{Interlayer \MR ~of \etcl ~at 13.5~kbar and 0.5~K. The
different lines correspond to different values of \thetat ~and
\phit. The (\thetat, \phit) values are, starting with the curve
having the highest resistance at 33~T; (23.7\deg, 160.9\deg);
(33.6\deg, 156.3\deg); (43.4\deg, 153.6\deg); (53.4\deg,
151.7\deg); (58.2\deg, 151.0\deg); (63.2\deg, 150.3\deg);
(68.2\deg, 149.7\deg); (73.6\deg, 149.08\deg); (78.1\deg,
148.6\deg) and (83.1\deg, 148.1\deg). Note that although the
\thetat ~values range widely, the \phit ~values vary only by about
12\deg. } \label{MR13kb}
\end{figure}
}

\def \fgfit{
\begin{figure}[t]
\centering
\includegraphics[height=9cm]{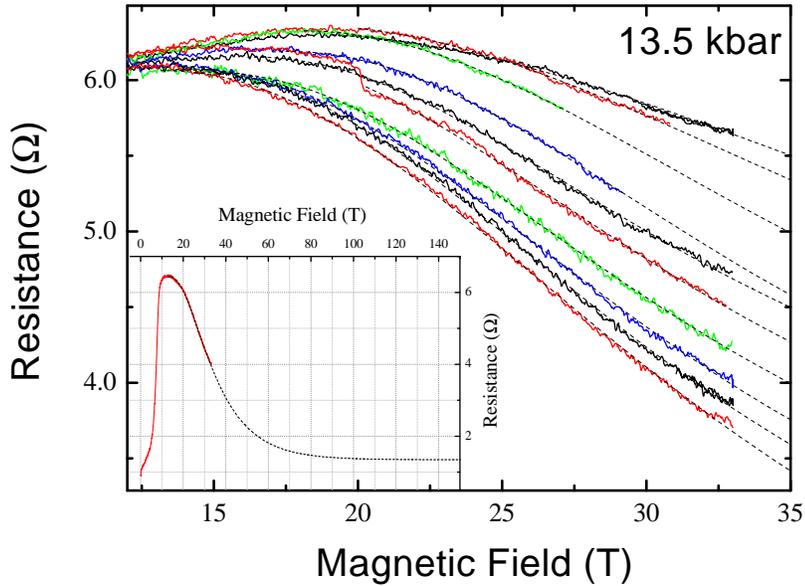}
\caption{The solid curves are the same data as in \fref{MR13kb}
and the dotted curves are fits using the form of equation
\eref{eqnlocal}. The inset shows the data for a field sweep at
0.5~K with $\theta_{\rm true}=89.8$\deg, and the corresponding
fitted line extrapolated to high magnetic fields. The
extrapolation of the Anderson localization model to very high
fields is shown for illustrative purposes only.} \label{fit}
\end{figure}
}

\def \fgmr14kb{
\begin{figure}[t]
\centering
\includegraphics[height=9cm]{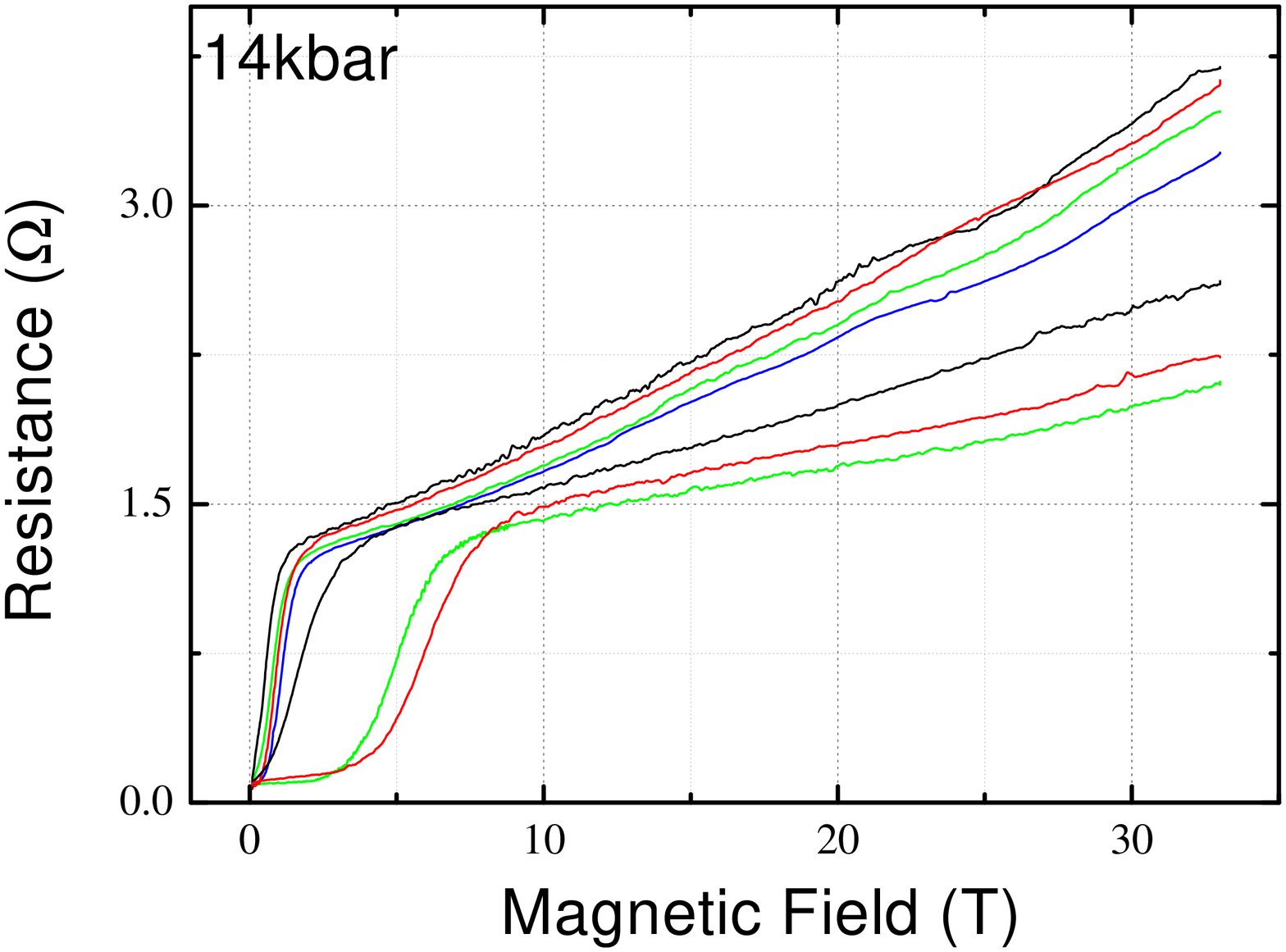}
\caption{Interlayer \MR ~of \etcl ~at 14.0~kbar and 0.5~K. The
different lines correspond to different values of \thetat ~and
\phit. The (\thetat, \phit) values are, starting with the curve
having the highest resistance at 33~T; (151.9\deg, 1.1\deg);
(141.9\deg, 358.5\deg); (26.6\deg, 352.0\deg); (25.6\deg,
349.8\deg); (104.1\deg, 354.0\deg); (89.2\deg, 352.9\deg);
(84.6\deg, 352.5\deg).} \label{MR14kb}
\end{figure}
}

\title[Superconductivity, incoherence and localization in \etcl]{Superconductivity, incoherence and Anderson
localization in the crystalline organic conductor \etcl ~at high
pressures}

\author{Paul Goddard\dag\footnote[4]{To whom correspondence should be addressed
(p.goddard@physics.ox.ac.uk)}, Stanley W. Tozer\ddag, John
Singleton\dag, Arzhang Ardavan\dag, Adam Abate\ddag~and
Mohamedally Kurmoo\S }

\address{\dag Clarendon Laboratory, Oxford University, Parks Road, Oxford, OX1 3PU, UK}

\address{\ddag National High Magnetic Field Laboratory, 1800 East Paul Dirac Drive, Tallahassee, Florida, FL 32310, USA}

\address{\S IPCMS, 23 rue du Loess, BP 20/CR, 67037 Strasbourg
Cedex, France}

\begin{abstract}
The conducting properties of the pressure-induced, layered organic
superconductor \etcl~have been studied at 13.5 and 14.0~kbar using
low temperatures, high magnetic fields and two-axis rotation. An
upper critical field that is significantly larger than that
expected from the Pauli paramagnetic limit is observed when the
field is applied parallel to the conducting layers. The angle
dependent magnetoresistance suggests incoherent transport between
the conducting layers at both pressures and the observed negative
magnetoresistance at 13.5~kbar can be explained by considering
Anderson localization within the layers. Further application of
pressure destroys the effects of localization.
\end{abstract}



\begin{sloppypar}
\section{Introduction}

The study of electrically conducting crystalline materials made
from organic molecules is an increasingly large and active area of
research. The high degree of anisotropy and the strong
electron-electron interactions in these crystalline organic
conductors means that they can possess a wide variety of ground
state phases~\cite{morrev}. Of particular interest are the class
of organic conductors which exhibit superconductivity, and the
similarities between these and the ``high-$T_{\rm c}$'' cuprate
superconductors are well documented~\cite{macca}.

One family of such organic conductors are made using the charge
transfer salts of the organic molecule
bis(ethlyene-dithio)tetrathiafulvalene (BEDT-TTF). These tend to
be layered materials with highly conducting planes of the BEDT-TTF
molecules separated by anion layers (for a complete overview
see~\cite{morrev,review}). Understanding the nature of the
interlayer transport in these materials is important in achieving
an grasp of the interactions that cause the superconductivity,
which in many cases is thought to be non BCS-like~\cite{chuck}. In
a pressure and temperature phase diagram the superconducting state
is frequently adjacent to a density-wave state caused by nesting
of quasi-one-dimensional (Q1D) pieces of the \fs, and this leads
to the suggestion that superconductivity is mediated by density
wave fluctuations. The topology of the \fs ~will obviously have an
effect on its nesting properties. Thus measurements that
investigate the nature of the interlayer transport in these
materials are of particular significance, as coherent transport
between the layers necessitates the existence of a three
dimensional, rather than two dimensional, \fs~\cite{squit}.
\fgphase

\etcl ~is so far unique in this family of organic crystal in that
each of the BEDT-TTF ions has an average charge of
$+\frac{2}{3}$e. Band calculations suggest that at room
temperature and ambient pressure this material should be a
semimetal~\cite{whang, mori}, and thermopower and conductivity
measurements confirm this~\cite{proc}. As the temperature is
reduced below $T\approx160$~K \etcl ~undergoes a
\mbox{(semi)metal-insulator} transition into a charge-density-wave
(CDW) state~\cite{struc}. A further \fs ~reconstruction occurs at
$T\approx60$~K and the sample then remains insulating down to very
low temperatures. The application of hydrostatic pressure
suppresses the onset of the density-wave state, and at 10.2 kbar a
superconducting state is formed with $T_{\rm c}$=4~K. In their
paper Lubczynski {\it et al.}~observe that the onset of
superconductivity coincides with a saturating magnetoresistance in
fields of up to 15 T and suggest that this implies that it is
quasi-two-dimensional (Q2D) carriers that are responsible for the
superconducting transport~\cite{lub}. Lubczynski {\it et al.} also
show that as the pressure is increased further the CDW transition
is completely suppressed causing the liberation of the Q1D
carriers and yielding a non-saturating \MR. Superposed over the
top of this \MR~they observe low-amplitude, low-frequency
Shubnikov-de Haas (SdH) oscillations, indicating that the Q2D
carriers are still present~\cite{lub}. Lubczynski {\it et al.} go
on to state that if the pressure is still further increased above
14 kbar the superconducting state is destroyed and the material
becomes metallic. \Fref{phase} shows the phase diagram proposed in
Reference~\cite{lub}.

\section{Experimental details}
\subsection{Sample properties}

\etcl ~crystallizes in the triclinic space group P\={1}, with
$a=11.214\pm0.002$~\AA, $b=13.894\pm0.002$~\AA,
$c=15.924\pm0.002$~\AA, $\alpha=94.74\pm1^{\circ}$,
$\beta=109.27\pm1^{\circ}$ and
$\gamma=97.03\pm1^{\circ}$~\cite{matt}. The conducting planes of
the (BEDT-TTF) molecules lie in the $ab$-plane with the long axis
of the molecules lying approximately parallel to the $c$-axis (see
Figure~\ref{struc}).

\fgstruc

Using their electronic band structure calculation Whangbo {\it et
al.} suggest that the in-plane \fs ~of \etcl ~at room temperature
and pressure consists of an elongated closed electron pocket
centred at the corner of the Brillouin zone, together with an open
electron \fs ~and an open hole \fs ~oriented along the $\Gamma$-Y
($b^*$) direction (see Figure~\ref{struc})~\cite{whang}. As
already mentioned, the material is a semimetal at room temperature
and the ratio of resistivities along the crystal axes $a$, $b$ and
$c$ is $\rho_a:\rho_b:\rho_c = 1:7:1000$~\cite{mori}.

The high purity, single crystal samples of \etcl ~are grown
electrochemically~\cite{matt} and are black platelets generally of
the order of $2\times2\times0.05$~mm$^3$ with the plane of the
plate corresponding to the highly conducting layers. The sample
used in this paper was cleaved so that it would fit in a pressure
cell and measured $120\times100\times45~\mu$m$^3$.

\subsection{Two-axis rotation at high hydrostatic pressure}

The experiments described in this paper involve rotation of the
\etcl ~sample about two axes in a high magnetic field at pressures
of $13.5\pm0.3$~kbar and $14.0\pm0.3$~kbar. These pressures were
achieved using the turnbuckle diamond anvil pressure cell (DAC)
shown in Figure~\ref{DAC}a. This DAC was designed and built in the
National High Magnetic Field Laboratory (NHMFL), Tallahassee and
consists of two natural diamonds enclosed in a body of
high-tensile BeCu~\cite{stan}. The highly polished, flat faces of
the diamonds are separated by a 65 $\mu$m thick, stainless steel
gasket coated in alumina. Two wires of pressed gold are attached
to each of the two broad faces of the sample using a paste made
from a mixture of graphite and gold. The sample is then placed
inside a cavity in the gasket, which is filled with a liquid
pressure medium, in this case glycerine. The DAC is then assembled
with the sample wires being electrically contacted to wires that
emerge from the cell body and the required room temperature
pressure is applied using a hydraulic press. Prior to removal from
the press the pressure can be maintained by means of a screw-type
locking mechanism. Figure~\ref{DAC}b shows the sample used in this
paper in position on the DAC, prior to assembly.

The magnetoresistance measurements were made using standard 4-wire
AC techniques ($f\sim80$~Hz) with the current applied in the
interplane direction ($I=1-25~\mu$A).

The pressure inside the sample cavity is calibrated by comparing
the fluorescence of a fragment of ruby located alongside the
sample in the DAC and the fluorescence of another fragment at the
same temperature, but ambient pressure~\cite{yen}. This is
measured using a helium-cadmium laser and a charge-coupled device
detector. The change in wavelength of the ruby R$_1$ fluorescence
line as a function of pressure is well known~\cite{eremets,yen}
and independent of temperature. Thus the pressure in the sample
cavity can be found at any temperature.

The two-axis rotation was achieved using an insert designed and
built in Oxford.
The angle between the normal to the $ab$-planes and the magnetic
field (the $\theta$-angle) can be varied continuously via a motor
and worm-drive arrangement and measured using a potentiometer. The
in-plane, or azimuthal angle (the $\phi$-angle) can be changed in
discrete steps using a retractable rod that travels along the axis
of the rotator. This should enable the B-field to be directed
along all possible sample directions.

All measurements were performed at NHMFL, Tallahassee at
temperatures of 500~mK and in fields of up to 33~T.
\figDAC
\subsection{Effect of sample misalignment}
\label{offset}
It is possible in experiments such as those presented here that
the sample axes are not accurately aligned, {\it i.e.} that the
normal to the $ab$-planes is inclined at an unknown angle to the
long axis of the DAC. This is especially true when the sample is
located within a pressure cell for a number of reasons. Probably
most important of these is the forces that act on the sample as
the temperature is reduced and the pressure medium freezes. The
necessity of having extremely thin contact wires means that the
sample will move easily under even small forces. Another origin of
this misalignment comes from the difficulty in cleaving these
samples to be so small. After cleaving it will not be possible to
guarantee that the plane of the sample will correspond exactly to
the $ab$-planes. Whatever the cause, this offset means that the
measured $\theta$ and $\phi$-angles are not the true angles
associated with the sample ($\theta_{\rm true}$~and~$\phi_{\rm
true}$). In fact by changing {\it either} of the measured angles
we will be changing {\it both} $\theta_{\rm true}$~{\it
and}~$\phi_{\rm true}$.

\fgaxis
In the general situation shown in \Fref{fgaxis}, we measure the
angles $\theta$ and $\phi$ (in the laboratory frame) and we wish
to know the angles $\theta_{\rm true}$ and $\phi_{\rm true}$ (in
the sample frame). The difference between these two sets of angles
is due to a misalignment of the two frames by an unknown angle in
an unknown direction such that the angle between the $z$-axis and
the $z'$-axis is $\epsilon$ and the angle between the $x$-axis and
the projection of the $x'$-axis onto the $xy$-plane is $\psi$.
Writing down the transformation matrix for converting laboratory
into sample frame leads us to the following equations for
$\theta_{\rm true}$~and~$\phi_{\rm true}$:
\trueangles \\
For a highly anisotropic, layered superconductor like \etcl ~the
upper critical field (\Bc2) will have a maximum when the magnetic
field is in the $ab$-planes \mbox{($\theta_{\rm
true}=90^{\circ}$)}~\cite{tinkham, nakamura}. Thus if we set the
magnetic field to be just below this maximum \Bc2, but above the
field required for zero resistance then we will observe a sharp
minimum in the resistance when we rotate the sample in a magnetic
field. The minimum in resistance will occur when the sample is
exactly in the plane of the sample \cite{loff}. Setting
$\theta_{\rm true}$ = 90$^\circ$ in~\Eref{cost} gives:
\vspace{0.15cm} \thetatrue90 \vspace{0.15cm}
Fitting this equation to a plot of the $\theta$ positions of the
resistance minima against $\phi$ yields values for the constants
$\epsilon$ and $\psi$. Substituting these values into Equations
(1), (2) and (3) gives a function for $\theta_{\rm true}$ that is
single-valued over the range \mbox{$0<\theta_{\rm
true}<180^{\circ}$}~and a function for $\phi_{\rm true}$~that is
single-valued over 360$^\circ$.

It is worth noting that as a result of this misalignment it will
not be possible to access the $\theta_{\rm true}=0$\deg~and
180\deg~directions for a given $\phi$~except when $\phi=\psi$.

\section{Results and discussion}

\subsection{Anisotropy of the upper critical field}

\figBc2

At both pressures measured in these studies the \etcl~sample
undergoes a transition into a superconducting state at low
temperatures. The $T_{\rm c}$'s at 13.5~kbar and 14.0~kbar are
found to be $3.2\pm0.5$~K and $2.8\pm0.5$~K respectively, where
$T_{\rm c}$ is defined by the midpoint of the resistive
transition. A study of the angular dependence of the upper
critical field (\Bc2) at each pressure was made by sweeping the
field at fixed \thetat-angles and using linear extrapolation to
extract \Bc2 as shown in Figure~\ref{figBc2}. For a discussion of
why this method of extracting \Bc2 is preferable to simply taking
the resistive midpoint, see Reference~\cite{chuck}, Section 4. The
results obtained by using this method are shown in
Figure~\ref{figGL}.

It was found that whilst \Bc2 varies strongly with \thetat~it
shows little or no \mbox{\phit-dependence}. It is possible to fit
the data with the functional form of the predictions of the
Ginsburg-Landau anisotropic effective mass
approximation~\cite{tinkham,nakamura,zuo};
\vspace{0.15cm} \GL \vspace{0.25cm}
in which \Bcperp ~is the upper critical field when $B$ is directed
along the interplane direction and $\gamma$ is the square root of
the ratio of the effective masses for interplane and in-plane
motion respectively.

Although it is seen from \Fref{figGL} that the data fit reasonably
well to the formula, it would not be correct to infer an
anisotropy of the effective masses from the values obtained from
the fit. This is because the Ginsburg-Landau theory is based on
the superconducting state being destroyed by orbital
effects~\cite{chuck}. In the case of highly anisotropic layered
materials such as \etcl, the flux lines can become trapped inside
the layers when a sufficiently high in-plane magnetic field is
applied, and in this situation the upper critical field in the
in-plane direction (\Bcpara) will become very high (for a more
complete discussion see Reference~\cite{chuck}). As \Bcpara~is not
particularly large in \etcl ~it is assumed that another effect is
limiting the superconductivity in the presence of low in-plane
fields.
\figGL

Despite this failure of Ginsburg-Landau theory to describe the
anisotropy of \Bc2, the functional form of~\Eref{eqnGL} is still
valid as it is just derived from a vector sum of two competing
critical fields and is independent of the mechanism responsible
for these critical fields~\cite{nam}. Thus the $\gamma$ factor
that is found from the fit is in fact just a measure of the
critical field anisotropy and is found to be $5.6\pm0.2$ and
$9.0\pm0.3$ for 13.5~kbar and 14.0~kbar respectively. In addition,
when the field is perpendicular to the conducting planes the
critical field will certainly be limited by orbital effects and so
we can use the fits in Figure~\ref{figGL} to obtain valid values
for \Bcperp~and hence the in-plane coherence length,
$\xi_\parallel$, using the relation \coh where $\Phi_0$ is the
flux quantum~\cite{zuo}. The values obtained, together with the
other results mentioned in this section are shown in
Table~\ref{Bcdata}.

One mechanism that might explain the origin of \Bcpara~in \etcl
~is the Pauli paramagnetic limit, also known as the
Clogston-Chandrasekhar limit~\cite{clogston,chandra}. In this case
the superconductivity is destroyed by Zeeman splitting of the
Cooper pairs at fields above $B_{\rm PPL}$. For an isotropic,
BCS-like superconductor \mbox{$B_{\rm PPL} = 1.84T_{\rm
c}$}~\cite{clogston}. Applying this to \etcl ~we find that $B_{\rm
PPL}= 5.9\pm0.9$~T and $5.2\pm0.9$~T for 13.5~kbar and 14.0~kbar
respectively. It is seen from Figure~\ref{figGL} and \Tref{Bcdata}
that the maximum \Bc2 are significantly larger than these values
for both pressures. It is apparent that this simple isotropic,
BCS-like analysis is not able to describe the in-plane upper
critical field in \etcl ~at these pressures and another mechanism
needs to be invoked in order to explain the data.

Similar results were found for the organic superconductors
(TMTSF)$_2$PF$_6$~\cite{lee} and \cuscn~\cite{loff,zuo}. In their
study Lee {\it et al.}~show that the superconducting state in
\pf6~at a certain pressure persists in applied in-plane fields of
up to 9~T~\cite{lee}, far exceeding the expected Pauli
paramagnetic limit, although in \pf6~the effect is more pronounced
than for \etcl ~due to the lower critical temperature ($T_{\rm
c}=1.2$~K). Lee {\it et al.}~also observe a marked
$\phi$-dependence of the in-plane upper critical field, which is
not observed in the material studied here. The conclusion of
Reference \cite{lee} is that \pf6~is a strong contender for
triplet Cooper pairing. In the case of \cuscn, Zuo {\it et
al.}~claim that the $B_{\rm PPL}$~derived from thermodynamic
arguments is much bigger than the value suggested by BCS
theory~\cite{zuo}, whereas Singleton {\it et al.} suggest that the
high in-plane critical field is caused by a field-induced
transition into a Fulde-Ferrell-Larkin-Ovchinnikov (FFLO)
superconducting state~\cite{loff}. The FFLO state occurs when
quasiparticles with opposite spin, whose \fs s are split by the
magnetic field, form Cooper pairs with non-zero
momentum~\cite{fflo}. These examples highlight some of the
possible mechanisms that enhance the in-plane critical field with
respect to $B_{\rm PPL}$ in organic superconductors, however none
of these mechanisms can be definitively attributed to
\etcl~without further extensive measurements.
\tabBcdata

\subsection{Angle dependence of the \MR}

\Fref{AMRO}a shows the interlayer \MR~of \etcl~as a function of
\thetat~in a steady magnetic field of 30~T. Three different values
of the measured $\phi$-angle are shown.
\figAMRO

As already mentioned in~\Sref{offset}, it is difficult in such a
measurement to separate the effects due to the azimuthal and
out-of-plane angles, because as the sample is rotated in the
\thetat-direction, \phit ~will also change to an extent determined
by the magnitude of the offset angle $\epsilon$. In light of this,
\Fref{AMRO}b shows the same curves as in \Fref{AMRO}a but as
function of \phit. Because of the mixing of angles it is not easy
to obtain a functional form for the \MR, but what is clear from
the Figure is that there is a minimum in resistance at
$\theta_{\rm true}=90$\deg~and a maximum when the field component
in the interlayer direction takes its maximum value. This
behaviour is contrary to the expectation that the \MR ~should be a
minimum when the magnetic field is parallel to the applied current
and a maximum when the field and current are perpendicular, as
expected from semiclassical transport theory~\cite{anm}. However
the situation is not without precedent in organic molecular
crystals.~\pf6~at 9.8~kbar~\cite{d&c,chkina},
(TMTSF)$_2$ClO$_4$~\cite{osada} and
$\tau$-[P-(S,S)-DMEDT-TTF]$_2$(AuBr)(AuBr)$_{\rm y}$~\cite{luis}
all show this type of angle dependence in the interlayer
resistance, and in all three cases it is attributed to the
interlayer transport being incoherent in nature, i.e. the \fs ~is
a two-dimensional object that exists only in the conducting layers
and the charge-carriers must undergo some kind of hopping
mechanism to move from layer to layer.

This situation can arise when an in-plane magnetic field is
applied to a sample even if the interlayer transport is coherent
to begin with. Increasing the in-plane component of the field
reduces the semiclassical width of the interplane component of the
quasiparticle orbits. At a certain field this width will become
less than the interlayer spacing, at which point the
quasiparticles are essentially confined to a single conducting
layer. This is the single-body confinement argument of Osada {\it
et al.}~\cite{osada} and their in-plane confinement field is given
by $B_{\rm conf}=4t_\perp/ev_{\rm F}l_c$, where $t_\perp$ is the
interlayer transfer integral, $v_{\rm F}$ is the Fermi velocity
and $l_c$ is the interlayer spacing. Typical values of $t_\perp$
and $v_{\rm F}$ for BEDT-TTF salts are 0.1~meV and 50~kms$^{-1}$
respectively~\cite{chuck,squit} and, for \etcl, $l_c=15.03$~\AA.
Hence $B_{\rm conf}\approx5$~T. For inclined fields it is
$B\sin\theta_{\rm true}$~that determines the in-plane component,
thus in fields of 30~T (as in \Fref{AMRO}) the interlayer
transport should be incoherent for $\theta_{\rm
true}\gtrsim10$\deg.

Strong and Clarke argue that the crossover from 3D to 2D transport
can occur at much lower values of the in-plane field
\cite{s&c,s&c&a}. Their reasoning is that the strong-correlation
effects in highly anisotropic materials can be enhanced by a small
in-plane field that introduces inelasticity into the interlayer
transport and effectively reduces the coherent interlayer transfer
integral to zero.

Whatever the cause of the incoherent interlayer transport, the
result is the same; the semiclassical transport theory breaks down
and the \MR ~will depend only on the interlayer component of the
magnetic field, $B\cos\theta_{\rm true}$. This would account for
the data of \Fref{AMRO}a.

\Fref{AMRO}c shows the \phit-dependence of the minimum in
resistance that occurs at $\theta_{\rm true}=90$\deg. The solid
line is a fit of the data to a cosine function with two-fold
symmetry. By considering the \fs ~shown in \Fref{struc} ~it is
seen that applying the magnetic field along the $\Gamma$-Y (or
$b^*$) direction will maximize the Lorentz force acting on both
the Q1D and Q2D carriers for in this direction the field is
perpendicular to the Q1D carrier velocity and also to the {\it
maximum} of the Q2D electron velocity. This is essentially the
same as the argument employed by Hussey {\it et al.} to describe
the ``High-$T_{\rm c}$'' compound YBa$_2$Cu$_4$O$_8$ in
Reference~\cite{hussey}. Here they conclude that the Q1D carriers
dominate the interlayer resistance. This may also be true in the
case of \etcl, but the possible presence of the elongated \fs
~pocket means that the contribution from the Q2D carriers cannot
be neglected. We can, however, conclude that the $b^*$ direction
corresponds to $\phi_{\rm true}=0$\deg.

\subsection{Magnetoresistance at 13.5 kbar}

\fgMR13kb

The \MR ~of \etcl ~at 13.5~kbar is shown in \Fref{MR13kb}. The
curves are data taken at different angles, the \thetat ~angle
varying from 23.7\deg ~to 83.1\deg ~and the \phit ~angle varying
very little. It is seen that for \thetat ~angles away from 90\deg
~the data in fields up to 15~T resemble the saturating \MR
~observed by Lubczynski {\it et al.}~\cite{lub}. This suggests
that we are in the region of the phase diagram proposed in that
paper where the CDW is not yet fully suppressed and the transport
is dominated by the Q2D carriers ({\it i.e.} in the
superconducting state, just below \Pc, see \Fref{phase}). At
fields greater than 15~T the \MR ~becomes negative. For each of
the curves \thetat ~is a constant and so the data in \Fref{MR13kb}
cannot be explained by the $B\cos\theta_{\rm true}$~dependence
derived from the incoherent interlayer transport models of Osada
or Strong and Clarke~\cite{osada,s&c,s&c&a}.

\subsubsection{Comparison with other materials.}

It is worthwhile making a diversion at this point to discuss the
results found in a few other organic conductors, as it may help to
shed some light on the situation in hand. It has already been
mentioned that at 9.8~kbar the behaviour of \pf6~can be explained
in terms of the Strong and Clarke many-body confinement
picture~\cite{s&c,s&c&a}. This leads to a decoupling of the
conducting layers and a $B\cos\theta$ dependence of the interlayer
\MR~\cite{d&c}. However at lower pressures (6-8.3~kbar) all
evidence for this decoupling is lost and the angle-dependence of
the interlayer \MR ~follows the semi-classical $B\sin\theta$
result~\cite{l&n}. This implies that the layers are coupled better
at lower pressures, which is contrary to the expectation that a
higher pressure will cause the electron orbitals to overlap to a
greater degree and thus {\it increase} the coherence in the
interlayer direction. Another curious matter concerns the \MR ~of
\pf6 at these low pressures when a in-plane field is applied
parallel to the Q1D \fs ~sheets. In this case the resistance
saturates in field of up to 7~T~\cite{l&n}. This contradicts the
predictions of {\it both} the semi-classical and the Strong and
Clarke theories, in which a non-saturating and diverging \MR ~are
expected respectively. This is itself similar to the results found
for (TMTSF)$_2$ClO$_4$, which is the material used to illustrate
Osada's single-body confinement effect~\cite{osada}. In the latter
material Naughton {\it et al.}~again noted a saturating \MR ~and
in fact used high fields to show that there is an onset of
negative \MR ~at around 28~T~\cite{ClO}. Another negative \MR ~is
recorded in the material
$\tau$-[P-(S,S)-DMEDT-TTF]$_2$(AuBr)(AuBr)$_{\rm y}$ of
Reference~\cite{luis}.

Thus it is seen that the effects observed in \etcl ~are not
unique. However this brings us no closer to an explanation. In the
much measured \pf6~theoretical considerations predict a field
induced re-entrance of the superconductivity~\cite{lebed}. However
in \etcl~there is no evidence to suggest that observed down-turn
of the \MR ~can be attibuted to a \mbox{re-entrance}. Moreover the
predicted restoration of the superconductivity in anisotropic
systems seems to require a purely Q1D system or a careful
alignment of the applied magnetic field parallel to the Q1D
\fs~sheets~\cite{lee,lebed,pf6}, whereas Q2D carriers have been
observed in \etcl~\cite{lub}, and it is seen from \Fref{MR13kb}
that the negative \MR~occurs for many different field directions.

\subsubsection{The effects of Anderson localization.}

The most likely mechanism responsible for the negative \MR ~at
high magnetic fields, at least in \etcl, is the presence of
Anderson localization within the conducting planes
\cite{fuku,kobay}. According to Fukuyama and Yoshida the
\mbox{variable-range} hopping mechanism responsible for conduction
in the localized regime would give rise to a large negative
\MR~\cite{fuku}. The explanation is that a random potential
produces the Anderson localized states {\it i.e.}~states below the
mobility edge, \Ec. If the Fermi energy, \Ef, lies below \Ec ~then
at low temperatures conduction is only possible by means of
variable-range hopping. Application of a magnetic field leads to a
Zeeman splitting of the electronic energy levels, and with further
increases in field the difference between \Ec ~and energy of an
electron with spin parallel to the magnetic field decreases,
leading to an increase in the conductivity. This effect will be
enhanced when \Ec-\Ef ~is small, such as in the vicinity of a
metal-nonmetal transition~\cite{kobay} (like that due to the
destruction of the CDW in \etcl). The conductivity resulting from
these considerations would have the form ~\cite{fuku};\\
\local
\\
Here $T_0$ is a characteristic temperature proportional to
(\Ef)$^d$, where $d$ is the dimensionality of the system, $n$ is
an integer (which for the simplest model is equal to
$d+1$~\cite{kobay}), $g$ is the effective g-factor (which is
assumed to be $\simeq2$), $\mu_{\rm B}$ is the Bohr magneton and
$\beta \simeq 1$. Following the analysis of Fukuyama and Yoshida
it is noted that because of the relationship between $d$ and $n$,
$\beta d/n$~is always of the order of 1, and equation
\Eref{eqnlocal} is not particularly sensitive to its precise
value. Thus we choose $\beta d/n = 1$ for a fit to the \MR ~data
of \Fref{MR13kb}. The result of the fitting is shown in \Fref{fit}
and it is seen that the fit is good at all angles. A value for
$T_0$ can be obtained from the temperature dependence of the
resistance, which follows the form $R\propto\exp[(T_0/T)^{1/n}]$
(although in \etcl ~this is complicated by the CDW and
superconducting transitions). Using this value, together with the
results obtained from the fit for data with an almost exactly
in-plane field, \Ec-\Ef ~is estimated to be about 6~meV. When the
Zeeman energy ($\frac{1}{2}g\mu_{\rm B}B$) of an electron
approaches \Ec-\Ef~the electrons will no longer be confined to the
localized states and the negative \MR~is expected to saturate.
This can be seen in the inset of \Fref{fit} which shows the fitted
\MR~saturating in a magnetic field of about 100~T which
corresponds to the estimated Zeeman energy of 6~meV.
\fgfit

Another prediction of the Anderson localization model of Fukuyama
and Yosida is the existence of a critical applied electric field,
proportional to $T^{4/3}$, above which the conduction can no
longer be considered as Ohmic~\cite{fuku}. At the low temperatures
used here the applied current produces an electric field across
the sample well below this critical value. However at higher
temperatures the resistance of the sample increases by several
orders of magnitude and this could lead to the critical electric
field being exceeded when even reasonably small currents are
applied. This might account for the region of non-Ohmic
conductivity observed by Lubczynski {\it et al.} between 150 and
30~K at ambient pressure.

If this Anderson localization effect is indeed responsible for the
observations in \etcl, then the mechanism responsible for
introducing disorder and randomizing the potential must be
considered. Ulmet {\it et al.} discuss the shape of the anion as a
possible source in the family of organic conductors (DMtTSF)$_2$X,
as they only observe the effects of localization in crystal with a
tetrahedral anion~\cite{ulmet}. However a much more likely
candidate in the case of \etcl ~is the CDW itself, which at
13.5~kbar has not yet been fully suppressed. The presence of the
density wave could introduce scattering centres in the conducting
planes and hence lead to localization. It is also possible that
even low levels of intrinsic disorder due to impurities or
vacancies could have a considerable effect on the conduction
properties of cystalline organic conductors of reduced
dimensionality. It has recently been suggested that the
anomalously broad superconducting transition in such crystals can
be accounted for by very small impurity concentrations
($\lesssim0.2\%$)~\cite{imp}.

\subsection{Magnetoresistance at 14.0 kbar}

\Fref{MR14kb} shows the \MR ~of \etcl ~at the slightly higher
pressure of 14.0~kbar. It is seen that the \MR ~is no longer
negative, and is in fact positive and non-saturating at all
angles. It is also seen that for field sweeps with \thetat~away
from 90\deg, low-amplitude, low-frequency SdH oscillations are
superposed over the background \MR. In the analysis of Lubczynski
{\it et al.}~this implies that the sample is in the region of the
phase diagram where the CDW has been completely suppressed and the
transport is dominated by the liberated Q1D carriers ({\it i.e.}
in the superconducting state, just above \Pc, see
\Fref{phase})~\cite{lub}. The background \MR ~for all angles can
be fitted well with a $B^{3/2}$ dependence, which is the result
found by Strong and Clarke~\cite{s&c&a} in \pf6~and described
using their incoherent transport model~\cite{s&c}.
\fgmr14kb

The fact that the negative \MR ~due to Anderson localization is
not observed at 14.0~kbar is unsurprising if we attribute the
random potential that causes the localization to the CDW
fluctuation. Here the CDW is no longer present and so localization
effects are not expected. Even if the Anderson localization is not
directly caused by the CDW we might expect the liberation of the
Q1D carriers to increase the carrier density to such an extent
that any scattering centres present are effectively screened from
the charge carriers and the variable-range hopping does not occur.
Thus at 14.0~kbar the angle-dependent and field-dependent \MR
~imply an interlayer incoherence in which Anderson localization is
no longer an important consideration.

\subsection{\Pc~and the superconducting state}

If the Anderson localization model suggested here is to be applied
to \etcl~it is necessary to reclarify the significance of the
critical pressure, \Pc. In Section 1 \Pc~was defined as the
pressure at which the CDW transition temperature approaches 0~K,
and hence at low temperatures it is also the pressure that
separates the regions of saturating and non-saturating \MR~in
fields of up to 15~T. Lubczynski {\it et al.}~attributed these
regions to the conduction being dominated by Q2D and Q1D carriers
respectively~\cite{lub}. However in the present analysis the
saturating magnetoresistance is merely a precursor to a negative
\MR~caused by Anderson localization. Thus it is not possible to
deduce the nature of the dominant carriers below \Pc, and the
dotted line in \Fref{phase}, which intercepts the pressure axis at
\Pc, merely separates the region where the effects of the
localization are observed and the region at higher pressure where
they are not. In the region above \Pc~the existence of the Q2D
carriers is implied by the SdH oscillations observed at high
fields.

It should be noted that \Pc~for the sample discussed in this paper
is approximately 1~kbar higher than the value obtained by
Lubczynski {\it et al.}~\cite{lub}. However if, as suggested, the
low temperature electronic properties in the region just below
\Pc~are dominated by random localization effects then a certain
degree of sample dependence is expected.

Lubczynski {\it et al.}~also suggest that the fact that the onset
of superconductivity coincides with the appearance of the
saturating magnetoresistance implies that it is Q2D carriers that
form the superconducting pairs. Extending this idea to the model
presented here suggests that there is a connection between the
superconducting state and the Anderson localization. This theory
is undermined by the fact that the superconductivity exists in the
region above \Pc~where the effects of localization are no longer
observed. In fact it seems much more likely that the onset of both
the superconductivity and the negative \MR~occurs in the region
close to the complete suppression of the CDW where there are just
enough free carriers available to observe any type of electrical
conduction effect whatsoever.

\section{Conclusions}

In summary, we have measured the angle-dependent interlayer \MR
~of the pressure-induced organic superconductor \etcl ~at two
pressures; one at which there is evidence for a CDW state, and one
at which this CDW is completely suppressed. At both pressures the
in-plane upper critical field exceeds the Pauli paramagnetic limit
and an incoherent interlayer transport mechanism is observed in
the presence of a in-plane magnetic field component. At the lower
pressure a negative \MR ~is seen, indicating that the transport is
dominated by variable-range hopping caused by a degree of Anderson
localization in the highly conducting organic layers. At the
higher pressure the \MR ~becomes positive and the effects of
Anderson localization are no longer observed following the
complete suppression of the CDW state. This implies a strong
connection between the CDW and the Anderson localization.

The Anderson localization model of Fukuyama and Yosida~\cite{fuku}
is currently the only mechanism that can explain all the effects
observed in \etcl.

We also note the similarities in physical properties of this
material and several other organic molecular crystals; in
particular $\tau$-[P-(S,S)-DMEDT-TTF]$_2$(AuBr)(AuBr)$_{\rm
y}$~\cite{luis}, (TMTSF)$_2$ClO$_4$~\cite{ClO} and
\pf6~\cite{l&n}; and suggest that it is possible that the Anderson
localization model described here could be used to explain the
negative and saturating magnetoresistances observed in these
materials.

This work is supported by EPSRC (U.K.). NHMFL is supported by the
U.S. Department of Energy (DoE), the National Science Foundation,
and the State of Florida. We thank Steve Blundell for stimulating
discussions.

\end{sloppypar}

\end{document}